\documentclass[epsfig,12pt]{article}

\usepackage{latexsym}
\usepackage{relsize}
\usepackage{geometry}
 \usepackage{amsmath, amssymb, amscd, xypic, graphicx}
\usepackage{slashed,color}
\usepackage{epstopdf}

\usepackage[colorlinks]{hyperref}
\usepackage{amsmath}
\usepackage[displaymath, tightpage]{preview}

\input xy
\xyoption{all}

 \usepackage{fancyhdr}                   
\def\beq{\begin{equation}}
\def\eeq{\end{equation}}
\def\beqn{\begin{eqnarray}}
\def\eeqn{\end{eqnarray}}

\newcommand{\ntwo}{${\mathcal N}=2\,$}

\newcommand{\cpn}{CP$(N-1)\,$}

\newcommand{\sma}{\left(\begin{smallmatrix}}
\newcommand{\smaa}{\end{smallmatrix}\right)}

\newcommand{\gsim}{\lower.7ex\hbox{$
\;\stackrel{\textstyle>}{\sim}\;$}}
\newcommand{\lsim}{\lower.7ex\hbox{$
\;\stackrel{\textstyle<}{\sim}\;$}}

\begin{document}

\begin{titlepage}

\begin{flushright}
FTPI-MINN-11/10, UMN-TH-2943/11\\
\today
\end{flushright}

\vspace{1cm}

\begin{center}
{  \Large \bf  \boldmath{$\mathcal N=(0,2)$} Supersymmetry and a Nonrenormalization Theorem }
\end{center}

\vspace{0.8cm}

\begin{center}
{\large
Xiaoyi Cui$^{\,a}$ and M.~Shifman$^{\,a,b}$}
\end {center}

\vspace{1mm}

\begin{center}

$^{a}${\it  Department of Physics, University of Minnesota,
Minneapolis, MN 55455, USA}\\[1mm]
$^b${\it  William I. Fine Theoretical Physics Institute,
University of Minnesota,
Minneapolis, MN 55455, USA}

\end{center}

\vspace{1cm}

\begin{center}
{\large\bf Abstract}
\end{center}

\vspace{2cm}
In this paper we continue the study of perturbative renormalizations in an $\mathcal{N}=(0,2)$ supersymmetric model. Previously we analyzed one-loop graphs in the heterotically deformed CP$(N-1)$ models. Now we extend the analysis of the $\beta$ function and appropriate $Z$ factors to two, and, in some instances, all loops in the limiting case $g^2\to 0$. The field contents of the model, as well as the heterotic coupling, remain the same, but the target space becomes flat. In this toy $\mathcal{N}=(0,2)$ model we construct supergraph formalism. We show, by explicit calculations up to two-loop order, that the $\beta$ function is one-loop-exact. We derive a nonrenormalization theorem valid to all orders. This nonrenormalization theorem is rather unusual since it refers to (formally) $D$ terms. It is based on the fact that supersymmetry combined with target space symmetries and ``flavor" symmetries is sufficient to guarantee the absence of loop corrections. We analyze the supercurrent supermultiplet (i.e., the hypercurrent) providing further evidence in favor of the absence of higher loops in the $\beta$ function.

\end{titlepage}

\newpage

\tableofcontents

\newpage

\section{Introduction}

In this paper we discuss multiloop calculations in a specific $\mathcal{N}=(0,2)$ linear sigma model. The motivation is two-folded. On the one hand, this is a continuation of our previous study \cite{CS1} of a class of two-dimensional $\mathcal{N}=(0,2)$ CP$(N-1)$ nonlinear sigma models (heterotic CP$(N-1)$ models for short). On the other hand, the linear model we suggest has its own field-theoretical significances, among which the most interesting are a peculiar 
supergraph technique and a version of nonrenormalization theorem. Surprisingly, it is a renoramlization theorem for $D$ terms! 

Two-dimensional  CP$(N-1)$ models  
emerged as effective low-energy theories on
the world sheet of non-Abelian strings in a class of 
four-dimensional \ntwo\, gauge theories~\cite{HT1,ABEKY,SYmon,HT2} (for reviews see  \cite{Trev}). 
Deforming these models in various ways (i.e. breaking supersymmetry down to ${\mathcal N}=1$)
one arrives at heterotically deformed \cpn models \cite{EdTo,SY1,BSY1,BSY2},
a very interesting and largely unexplored class of models characterized by two coupling constants:
the original asymptotically free coupling and an extra one describing the strength of the heterotic deformation.
These two-dimensional models exhibit highly nontrivial dynamics, with a number of phase transitions. This fact was 
recently revealed \cite{largen} in the
large-$N$ solution of the model.

Our task is the study of perturbation theory in two-dimensional heterotic models. Many general aspects of $\mathcal{N}=(0,2)$ models in perturbation theory were discussed in \cite{Witten05, hep-th/0604179}. 
The problem we address is more concrete.
In \cite{CS1} we studied particular renormalization properties and calculated the one-loop $\beta$ functions in the CP$(N-1)$ models heterotically deformed in a special way.
Written in components\footnote{The superfield expression for the heterotically deformed CP$(N)$ models can be found e.g. in \cite{CS1}.},
 the Lagrangian of the heterotic CP$(1)$ model takes the following form \cite{SY1}:
 \beq
 \mathcal{L}_{\rm het} = \mathcal{L}_{(2,2)}+\mathcal{L}_{(0,2)}\,,
 \eeq
 where
 \beq
 {\mathcal{L}}_{(2,2)}
=
 G\left\{
\partial^\mu\phi\partial_\mu\phi^\dagger+i\bar{\psi}\slashed{\partial}\psi -
{2i}  \,\frac{1}{\chi}\, \bar{\psi}\gamma^\mu\psi\,\phi^\dagger\partial_\mu\phi-\frac{2}{\chi^2}\,\psi_L^\dagger\psi_L\,\psi_R^\dagger\psi_R 
\right\}\,,
\label{N22model}
 \eeq
 and\,\footnote{The sign in front of the term $\zeta_R^\dagger\, \zeta_R \psi_L^\dagger\psi_L$ in (\ref{N02model})
 is opposite to that in \cite{SY1} due to a typo in \cite{SY1}. Also notice that the definition of $\gamma$ in this paper corresponds to $\gamma g^2$ in \cite{SY1}.
 The reason for rescaling of the deformation parameter 
 compared to \cite{SY1} is that both $g^2$ and $|\gamma |^2$ here are genuine loop expansion parameters, as the reader will see later. }
\beqn
&&
{\mathcal L}_{(0,2)}= 
\zeta_R^\dagger \, i\partial_L \, \zeta_R  + 
\left[\frac{\gamma}{g^2} \, \zeta_R  \,R\,  \big( i\,\partial_{L}\phi^{\dagger} \big)\psi_R
+{\rm {\rm H.c.}}\right] 
\nonumber
\\[3mm]
&&
+\frac{|\gamma |^2}{g^2} \left(\zeta_R^\dagger\, \zeta_R
\right)\left(R\,  \psi_L^\dagger\psi_L\right)
+
G\, \left\{
\frac{2 |\gamma |^2}{g^2\chi^2}\,\psi_L^\dagger\,\psi_L \,\psi_R^\dagger\,\psi_R
\right\}\,.
\label{N02model}
\eeqn
We denote by $G$
the K\"ahler  metric on the target space, 
\beq
G =
\frac{2}{g^2\,\chi^{2}}\,,
\label{kalme}
\eeq
$R$ is the Ricci tensor,
\beq
 R =\frac{2}{\chi^2}\,,
\label{Atwo}
\eeq
and we use the notation 
\beq
\chi \equiv 1+\phi\,\phi^\dagger\,.
\label{chidef}
\eeq
The coupling $g^2$ enters through the metric, while the deformation coupling $\gamma$
appears in Eq.~(\ref{N02model}). 
In the previous paper \cite{CS1} we determined the one-loop $\beta$ functions
\beqn
\beta(g^2) &\equiv& \frac \partial{\partial\text{ln}\mu}g^2(\mu)
=
-\frac{g^4}{2\pi} + ...\,,
\label{betafsp}
\\[2mm]
\beta(\gamma) &\equiv& \frac \partial{\partial\text{ln}\mu}\gamma(\mu)
=
\frac{\gamma}{2\pi}\left(\gamma^2-g^2\right) + ...\,,
\label{betafs}
\eeqn
where the dots stand for two-loop and higher-order terms. The heterotic deformation does not
affect $\beta(g^2)$ which stays the same as in the ${\mathcal N}=(2,2)$ CP(1) model.
Among other results, we calculated the law of running of the ratio $\rho= {\gamma^2}/{g^2}$. If in the ultraviolet (UV) limit $\rho$ is chosen to be smaller than $1/2$, in the infrared (IR) it runs to $\rho\to 1/2$, which is the fixed point for this parameter. With $\rho\leq1/2$ in the UV, the theory is asymptotically free.

Now we undertake the next step: multiloop graphs. However, this is not easy. At two and higher loops interplay between $g^2$ and $\gamma$ is contrived.  The impact of the deformation term was not studied before. It seems reasonable to start from untangling $\gamma$ from the nonlinear target space. In the heterotic CP$(1)$ model there is a fermion flavor symmetry. From a practical point of view, we want to understand this symmetry by probing it in a simpler setup. So we 
will focus on a simpler, linear version of the $\mathcal{N}=(0,2)$ sigma model, 
(setting $g^2 =0$) which serves our purposes at this stage.

We start from developing an appropriate $\mathcal{N}=(0,2)$ supergraph technique to carry our an explicit two-loop calculation. The result is as follows: the interaction term proportional to $\gamma$ is not renormalized, and so are the $Z$ factors of the superfield $A$ (see Eq.~(\ref{16})). The $Z$ factors of the superfields $\mathcal{B}$ and $B$ are renormalized, but this is just an iteration of the one-loop contribution. Then we prove the nonrenormalization theorem, which extends the first result to all orders. What is remarkable is the fact that the nonrenormalization theorem emerges for a $D$ term provided there are certain target space conditions.
 Thus, up to two-loop order, the $\beta$ function in the heterotic model at hand is 
 \beq
 \beta(\gamma) = \frac {\gamma^3} {2\pi}\,.
 \label{simplifiedbeta}
 \eeq
 This is compatible with (\ref{betafs}), of course.
 Due to the fact that the nonrenormalization theorem generally fails to detect the geometric progression in 
 the $Z$ factors of $\mathcal{B}$ and $B$, at the moment we can not directly extend this 
 result to three loops and higher in $\beta(\gamma)$. But it is reasonable to conjecture that this is the case. An argument substantiating this statement is presented in Sec.~\ref{sec7}.

The paper is organized as follow. In Sec.~\ref{sec2} we introduce the simplified heterotic $\mathcal{N}=(0,2)$ linear model, which captures in full the quantum behavior of the deformation strength $\gamma$. In Sec.~\ref{sec3} we give the Feynman rules for supergraph calculations in $\mathcal{N}=(0,2)$ theories. In Sec.~\ref{sec4}, we calculate the two-loop contribution
to  $\beta(\gamma)$. Vanishing of certain diagrams provides us with an indication of the nonrenormalization theorem. In Sec.~\ref{sec5}, we give the $D$ term nonrenormalization theorem, which is valid perturbatively. In Sec.~\ref{sec6} we extend this
 statement beyond perturbation theory. In Sec.~\ref{sec7} we analyze the supercurrent supermultiplet of this model (the so-called hypercurrent), following the line of reasoning of \cite{DS}.

\section{An \boldmath{$\mathcal{N}=(0,2)$} linear model}
\label{sec2}

In our previous work \cite{CS1} we showed that in the CP$(1)$ model, there is a fermionic SU$(2)$ flavor symmetry, which mixes the chiral fields $\mathcal{B}$ and $B$ (see Eq.~(\ref{fsu2})). To mimic this phenomenon, we introduce a simplified $\mathcal{N}=(0,2)$ linear model, which emphasizes the mechanism of the $\mathcal{N}=(2,2)$ deformation and retains the fermion flavor symmetry. 

We begin by briefly reviewing  $\mathcal{N}=(0,2)$ supersmmetry and  some notations.
We define the left moving and right moving derivatives as
\beq
\partial_L=\partial_t+\partial_z\,,\qquad \partial_R=\partial_t-\partial_z\,,
\eeq 
and use the following definition for the superderivatives:
\beq
D_R = \frac \partial {\partial \theta_R}-i\theta_R^\dagger \partial_L\,,\qquad
\bar D_R = -\frac\partial{\partial \theta_R^\dagger}+i\theta_R\partial_L\,.
\eeq
Their commutator gives $\{D_R,\bar D_R\} = 2i\partial_L\,$, as it should. All integrations and differentiations
 are understood as acting from the left, if not stated to the contrary.
The shifted space-time coordinates that satisfy the chiral condition are 
\beq
y^0=t+i\theta_R^\dagger\theta_R\,,\qquad y^1=z+i\theta_R^\dagger\theta_R\,.
\label{defy}
\eeq 
The antichiral counterparts are
\beq
\tilde y^0=t-i\theta_R^\dagger\theta_R\,,\qquad \tilde y^1=z-i\theta_R^\dagger\theta_R\,.
\label{defyt}
\eeq 
Under supersymmetric transformation $ \delta_\epsilon+\delta_{\bar \epsilon}$
\beqn
 \theta_R \to \theta_R+ \epsilon\,,&\qquad&  \theta^\dagger_R \to \theta^\dagger_R+ \bar\epsilon\,,\nonumber\\[3mm]
 y^\mu \to y^\mu+ 2i\bar\epsilon\theta_R\,,&\qquad&  \tilde y^\mu \to \tilde y^\mu -2i\theta^\dagger_R\epsilon\,,
\eeqn
where $\mu = 0,1$.

We can now define the chiral $\mathcal{N}=(0,2)$ superfields in our model,
\beqn
&&
A(y^\mu,\theta_R)=\phi(y^\mu)+\sqrt{2}\theta_R\psi_L(y^\mu)\,,\nonumber\\[3mm]
&&
B(y^\mu,\theta_R)=\psi_R(y^\mu)+\sqrt{2}\theta_R F(y^\mu)\,,\nonumber\\[3mm]
&&
\mathcal B(y^\mu,\theta_R)=\zeta_R(y^\mu)+\sqrt{2}\theta_R \mathcal F(y^\mu)\,.
\label{fieldsdef}
\eeqn
Here $\phi$, $\psi_L$, $\psi_R$ and $\zeta_R$ describe physical degrees of  freedom, while $F$ and $\mathcal{F}$ will enter without derivatives and, thus, can be eliminated by virtue of equations of motion. 

In the $\mathcal N=(0,2)$ superfield formalism the Lagrangian of the simplified model is as follow:
\beq
\mathcal{L}=\frac12\int d^2\theta_R\,\left[  \frac12\left( iA^\dagger \partial_R A-iA\partial_R A^\dagger \right) + B^\dagger B+\mathcal{B}^\dagger \mathcal{B} -\left(\gamma\mathcal{B}BA^\dagger + {\rm H.c.}\right)\right]\,.
\label{16}
\eeq
In the component language, after eliminating $F$ and $\mathcal{F}$, we have 
\beqn
\mathcal{L}&=&\partial^\mu\phi^\dagger\partial_\mu \phi + i\bar \psi\slashed{\partial} \psi +i\zeta^\dagger_R \partial_L\zeta_R + \left[\gamma \zeta_R\psi_R\partial_L\phi^\dagger+{\rm H.c.}\right]\nonumber\\[3mm]
&&+\gamma^2\left(\zeta^\dagger_R\zeta_R \right)\left(\psi^\dagger_L\psi_L \right)+\gamma^2\left(\psi^\dagger_R\psi_R \right)\left(\psi^\dagger_L\psi_L \right)\,.
\label{linearL}
\eeqn
Note that $\mathcal{N}=(0,2)$ supersymmetry completely fixes the second line in terms of the first line.

The Lagrangian is invariant under SU$(2)$ rotations of $B$ and $\mathcal{B}$. Actually,
 if we define an SU$(2)$ superfield doublet
\beq
\Psi = \sma B\\ \mathcal{B}\smaa\,,
\eeq
the part of the Lagrangian that involves all right-handed fermions can be rewritten as 
\beq
\frac12\int d^2\theta_R\, \Psi^\dagger_a\Psi_a+\left[\frac\gamma 2 A^\dagger \varepsilon^{ab}\Psi_a\Psi_b+{\rm H.c.}\right]\,,
\label{fsu2}
\eeq
which is obviously SU$(2)$ invariant.

Comparing with Eq.~(\ref{N02model}), we indeed see that Eq.~(\ref{linearL}) is the limiting case of the former with 
${\gamma^2}/{g^2}\to \infty$. The opposite limiting case, ${\gamma^2}/{g^2}\to 0$, is well-understood; it is just 
the undeformed 
${\mathcal N} =(2,2)$ model in Eq.~(\ref{N22model}). The model in Eq.~(\ref{linearL}) 
can be viewed as a preparatory step to developing perturbation theory in the $\mathcal{N}=(0,2)$ heterotic CP$(N-1)$ models. We will show that this model exhibits a nonrenormalization theorem. The proof of the latter strengthens our understanding of heterotic supersymmetry.

\section{Supergraph method}
\label{sec3}

In this section we explicitly formulate superfield/supergraph calculus for the 
given model. Calculations in the
 $\mathcal{N}=(0,1)$ language were previously discussed in the literature, 
 see e.g. \cite{GGMT, LO}. We feel that it is worth developing a similar formalism for $\mathcal{N}=(0,2)$ theories, for the following reasons. First, most $\mathcal{N}=(0,2)$ models  can be obtained as deformations from $\mathcal{N}=(2,2)$, where holomorphic structures are crucial. It would be best if we preserve them explicitly. Second, this language is useful in deriving the nonrenormalization theorem of Sec.~\ref{sec5}, a phenomenon not so easy to see when manipulating with $\mathcal{N}=(0,1)$ superalgebras. Third, so far no calculations were performed at two-loop level. The tools we develop here are expected to be 
 helpful in the heterotic CP$(N-1)$ models too.

To derive the superpropagator, we define the functional variation for a bosonic chiral and antichiral superfields,
\beqn
\frac \delta{\delta A(y,\theta_R)} A'(y', \theta'_R) &=& \delta(y-y')\delta(\theta_R-\theta'_R)\,,\nonumber\\ [3mm]
\frac \delta{\delta A^\dagger(\tilde y,\theta^\dagger_R)} A'^\dagger(\tilde y', \theta'^\dagger_R) &=& \delta(\tilde y-\tilde y')\delta(\theta_R^\dagger-\theta'^\dagger_R)\,,
\eeqn
where $y$ and $\tilde y$ are defined in Eq.~(\ref{defy}) and (\ref{defyt}).
For a generic function $F(x,\theta_R,\theta_R^\dagger)$, we have
\beqn
&&\frac \delta{\delta A(y,\theta_R)}\int d^2 x' d\theta'_R d\theta'^\dagger_R\, A(y', \theta'_R) F(x',\theta'_R,\theta'^\dagger_R) \nonumber\\[3mm]
&=& \int d^2 y' d\theta'_R d\theta'^\dagger_R\, \delta(y-y')\delta(\theta_R-\theta'_R) F(y'-i\theta'^\dagger_R\theta'_R,\theta'_R,\theta'^\dagger_R)\nonumber\\[3mm]
&=&-\int d\theta_R^\dagger\, F(y-i\theta^\dagger_R\theta_R,\theta_R,\theta^\dagger_R) = \bar D_R F(x,\theta_R,\theta^\dagger_R)\,.
\eeqn

Similarly,
\beqn
&&\frac \delta{\delta A^\dagger (\tilde y,\theta_R^\dagger)}\int d^2 x' d\theta'_R d\theta'^\dagger_R\,  F(x',\theta'_R,\theta'^\dagger_R) A^\dagger(\tilde y', \theta'^\dagger_R)\nonumber\\[3mm]
&=& \int d^2 y' d\theta'_R d\theta'^\dagger_R\, F(\tilde y'+i\theta'^\dagger_R\theta'_R,\theta'_R,\theta'^\dagger_R) \delta(\tilde y-\tilde y')\delta(\theta^\dagger_R-\theta'^\dagger_R)\nonumber\\[3mm]
&=&\int F(\tilde y+i\theta^\dagger_R\theta_R,\theta_R,\theta^\dagger_R)\,d\theta_R =  F(x,\theta_R,\theta^\dagger_R)\overleftarrow{D}_R\,.
\eeqn
Note that we intentionally write $D_R$ acting from the right, because we want our expression to be 
explicitly Hermitean-conjugate to the previous result. 

On the other hand, we compare the result with 
\beqn
&&\int d^2 x' d\theta'_R d\theta'^\dagger_R\, F(x',\theta'_R,\theta'^\dagger_R) \bar D_R \delta(x-x')\delta(\theta_R^\dagger-\theta'^\dagger_R)\delta(\theta_R-\theta'_R)\nonumber\\[3mm] 
&=& -\int d^2 x' d\theta'_R d\theta'^\dagger_R\, \delta(x-x')\delta(\theta_R^\dagger-\theta'^\dagger_R)\delta(\theta_R-\theta'_R) \bar D_R F(x',\theta'_R,\theta'^\dagger_R) \nonumber\\[3mm]
&=& -\bar D_RF(x,\theta_R,\theta^\dagger_R)\,,
\eeqn
which implies that, upon integration,
\beqn
\frac \delta{\delta A(x,\theta_R,\theta^\dagger_R)} A'(x', \theta'_R,\theta'^\dagger_R) &=& -\bar D_R \delta(x-x')\delta(\theta_R^\dagger-\theta'^\dagger_R)\delta(\theta_R-\theta'_R)\,,\nonumber\\ [3mm]
\frac \delta{\delta A^\dagger(x,\theta_R,\theta^\dagger_R)} A'^\dagger(x', \theta'_R,\theta'^\dagger_R) &=& - \delta(x-x')\delta(\theta_R^\dagger-\theta'^\dagger_R)\delta(\theta_R-\theta'_R)\overleftarrow{D}_R\,.
\eeqn
For a chiral field $J_A$, we have the projection 
\beq
\frac{\bar D_R D_R} {2i\partial_L} J_A = \frac{\{\bar D_R, D_R\}} {2i\partial_L} J_A = J_A\,.
\eeq
Using this we can conveniently pass from the $F$ term to the integration over the full superspace, namely
\beqn
&&\int d^2 x d\theta_R\, A J_A 
= \int d^3z\, A\frac{-D_R}{2i\partial_L} J_A = \int d^3z\, J_A \frac{D_R}{2i\partial_L} A\,,\nonumber\\[3mm]
&&\int d^2 x \, J^\dagger_A A^\dagger d\theta^\dagger_R
= \int d^3z\, A^\dagger \frac{\bar D_R}{2i\partial_L} J^\dagger_A = \int d^3z\, J^\dagger_A \frac{- \bar D_R}{2i\partial_L} A^\dagger\,.
\eeqn
Here and in what follows in this section we use $z$ to denote the triplet 
of (super)coordinates $(x^\mu, \theta_R, \theta^\dagger_R)$.
Note that the currents $J_A$ and $J^\dagger_A$ are Grassmannian.
We can write the partition function as
\beq
Z[J_A, J_A^\dagger ] = \int \mathcal{D} A\mathcal{D}A^\dagger
\exp \left(  i\int d^3 z\, \frac i2 A^\dagger \overleftrightarrow{\partial_R} A+ A\frac{-D_R}{2i\partial_L} J_A + A^\dagger \frac{\bar D_R}{2i\partial_L} J^\dagger_A\right)\,,
\eeq
 and, by virtue of the functional integration, we get
 \beqn
 && \exp\left[-\frac i2\int d^3 z \sma J_A & J^\dagger_A\smaa
 \sma \frac{D_R}{2i\partial_L} & 0\\ 0 &\frac{-\bar{D_R}}{2i\partial_L} \smaa
 \sma 0 & -\frac{2i}{\partial_R} \\[1mm]
 \frac{2i}{\partial_R} & 0 \smaa
 \sma \frac{-D_R}{2i\partial_L} J_A\\ \frac{\bar D_R}{2i\partial_L} J^\dagger_A\smaa \right] \nonumber\\[3mm]
 &=& \exp  \int d^3 z -\frac i2\left(J^\dagger_A \frac {1}{\Box} J_A +J_A \frac{-1}{\Box} J^\dagger_A \right)\,.
 \eeqn
 As a result we get the Feynman propagator for the chiral field $A$ in the form 
 \beq
 \langle 0| T \{A(x,\theta_R, \theta^\dagger_R)\,,\, A^\dagger(y,\eta_R, \eta^\dagger_R)\}|0\rangle = \frac{i}{\Box}\delta(x-y)\delta(\theta^\dagger_R - \eta^\dagger_R)\delta(\theta_R-\eta_R)\,.
 \eeq
 
Using the same line of reasoning now we will determine the propagators for 
the superfields 
$B$ and $\mathcal{B}$. Note that due to the fermionic symmetry (see Eq.~(\ref{fsu2})), they are exactly the same. Take $B$ for example; 
the partition function is 
 \beq
Z[J_B, J_B^\dagger ] = \int \mathcal{D} B\mathcal{D}B^\dagger
\exp\left(i\int d^3 z \frac 12 B^\dagger B+ B\frac{D_R}{2i\partial_L} J_B + B^\dagger \frac{-\bar D_R}{2i\partial_L} J^\dagger_B\right)\,.
\eeq
 By virtue of the functional integration, we arrive at
 \beqn
 && \exp\left[-\frac i2\int dz \sma J_B & J^\dagger_B\smaa
 \sma \frac{-D_R}{2i\partial_L} & 0\\ 0 &\frac{\bar{D}_R}{2i\partial_L} \smaa
 \sma 0 & -2 \\[2mm]
2 & 0 \smaa
 \sma \frac{D_R}{2i\partial_L} J_B\\ \frac{-\bar D_R}{2i\partial_L} J^\dagger_B\smaa \right] \nonumber\\[3mm]
 &=& \exp \int d^3 z -\frac i2\left(J^\dagger_B \frac {1}{i\partial_L} J_B +J_B \frac{-1}{i\partial_L} J^\dagger_B \right)\,.
 \eeqn
 As a result,
 \beq
 \langle 0| T \{B(x,\theta_R, \theta^\dagger_R) \,,\,B^\dagger(y,\eta_R, \eta^\dagger_R)\}|0\rangle = \frac{-1}{\partial_L}\delta(x-y)\delta(\theta^\dagger_R - \eta^\dagger_R)\delta(\theta_R-\eta_R)\,.
 \eeq
 The same applies to $\mathcal{B}$.
 
 Now, let us pass to the interaction vertices. They can be obtained by considering 
 \beqn
&& S_{\rm int}\left[\frac\delta{\delta J}\right] J_{\mathcal{B}} J_B J^\dagger_{A} = 
 -\frac \gamma 2 \int d^3z \left(\frac \delta{\delta J_\mathcal{B}} \frac \delta{\delta J_B} \frac \delta{\delta J^\dagger_A}\right) J_\mathcal{B}(z_1) J_B(z_2) J^\dagger_A(z_3)\nonumber\\[3mm]
 &=&\frac \gamma2 \int d^3 z \bar D_R \delta (z-z_1)\, \bar D_R (z-z_2)\, \delta(z-z_3)\overleftarrow{D}_R\,.
 \eeqn
 
\begin{figure}
 \begin{center}
 \includegraphics[width = 6.05in]{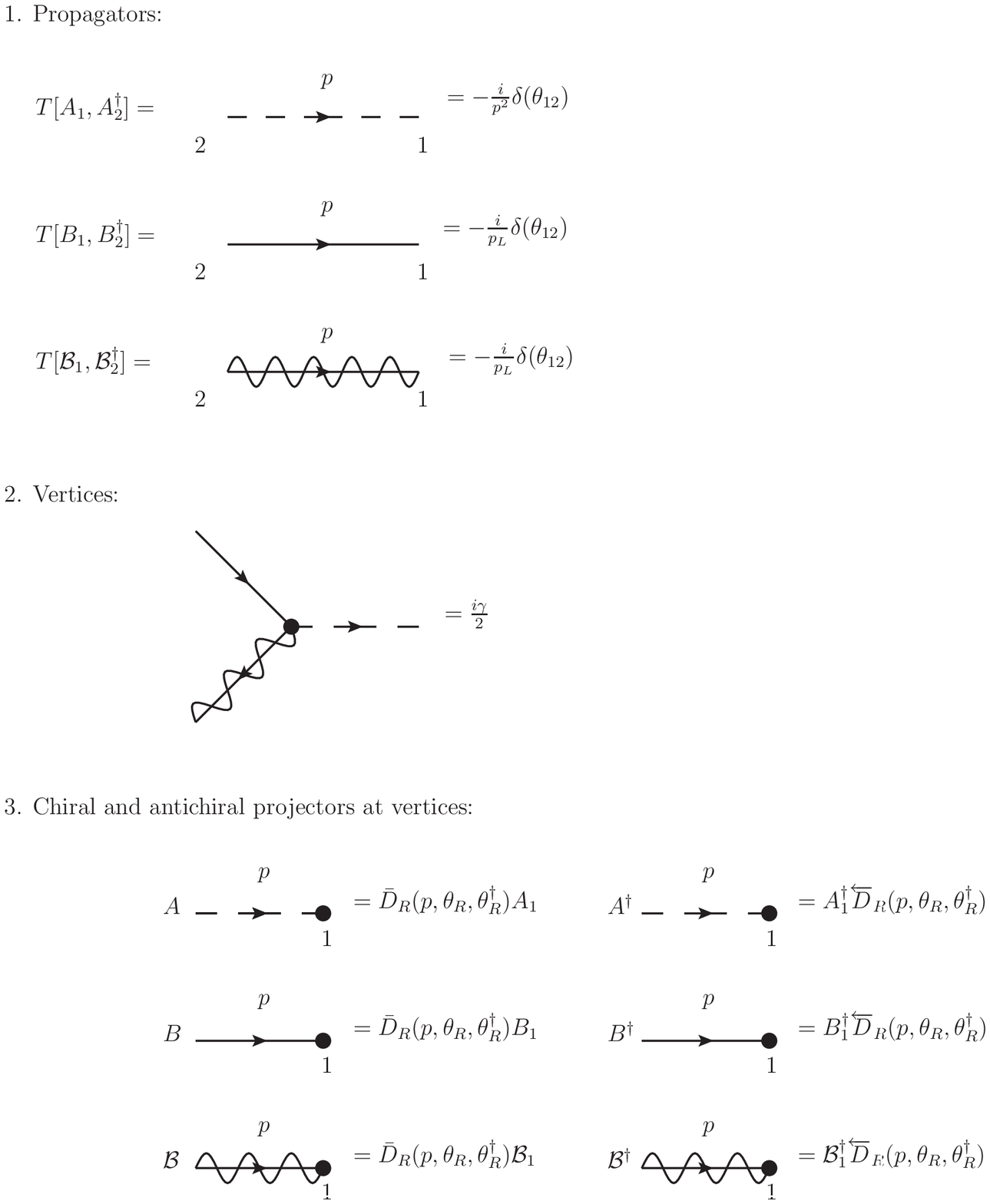}
 \end{center}
 \caption{\small  Feynman rules for the linear $\mathcal{N}=(0,2)$ sigma model.}
 \label{feynrule}
 \end{figure}

 We can summarize the Feynman rules for the model at hand in the momentum space:
 \begin{itemize}
 \item For each propagator $\langle0|T \{A_1\,,\, A^\dagger_{2}\}|0\rangle$, write $$-\frac i {p^2} \delta(\theta_{12})$$ where $\delta(\theta_{12}) = \delta(\theta^\dagger_1-\theta^\dagger_2)\delta(\theta_1-\theta_2)$; for each propagator $\langle0|T \{B_1\,,\, B^\dagger_2\}|0\rangle$ or $\langle0|T\{ \mathcal{B}_1\,,\, \mathcal{B}^\dagger_2\}|0\rangle$, write 
 $$-\frac i {p_L} \delta(\theta_{12})\,,$$
 with the momentum $p$ flowing from $2$ to $1$.
 \item For each vertex, write $i\frac\gamma2$. 
 \item For each propagator that connects 
 a chiral field to the vertex, put 
 $\bar D_R (p,\theta^\dagger_R,\theta_R)$ acting on it; for that connecting 
 an antichiral field, put $\overleftarrow{D}_R(p,\theta^\dagger_R,\theta_R)$ acting on it, 
 where $p$ is the momentum that flows into the vertex through the propagator.
 \item Integrate over $\int d^2 \theta_R$ and impose momentum conservation at each vertex, integrate over the momentum $\int \frac {d^2p}{(2\pi)^2}$ for each loop.
 \item For each external chiral or antichiral line, we have a factor for the field, but no $D_R$ or $\bar D_R$ factors. 
 \end{itemize}
 This set of the Feynman rules is displayed in Fig.~\ref{feynrule}.
 
 To facilitate our calculation, let us present here some useful identities. Verification of these 
 identities is straightforward and is left as an exercise for the reader. In what follows, we will omit the subscript $R$ in $\theta_R$ and $D_R$,
 \beqn
 \delta(\theta_1 - \theta_2)\overleftarrow{D}_2(\theta_2, p)&=& -D_1(\theta_1, -p)\delta(\theta_1-\theta_2)\,,\nonumber\\ [3mm]
 \bar D_1 D_1(\theta_1, p)\delta(\theta_{12})|_{\theta_1 = \theta_2}&=&  - D_2\bar D_2(\theta_2, -p)\delta(\theta_{12})|_{\theta_1 = \theta_2} = 1\,.
  \eeqn

\section{One and two-loop results}
\label{sec4}

\begin{figure}
\begin{center}
\includegraphics[width=5in]{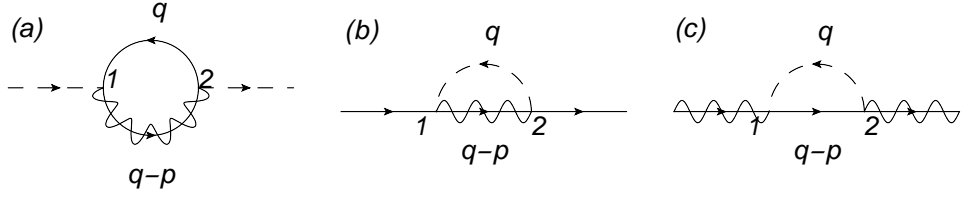}
\end{center}
\caption{\small We use dashed line for 
the field $A$, straight arrowed line for the field $B$, and straight  with wavy lines superimposed for the field $\mathcal{B}$.}
\label{1loop_110324}
\end{figure}

Now we are ready to undertake the loop calculations using the superfield technique. 
We start from the Lagrangian with the bare coupling in UV, and evolve it down, where we have
\beq
\mathcal{L}=\frac12\int d^2\theta_R\, \frac12 Z_A\left( iA^\dagger \partial_R A-iA\partial_R A^\dagger \right) + Z_B B^\dagger B+Z_\mathcal{B} \mathcal{B}^\dagger \mathcal{B} -Z_\gamma\left(\gamma_0\mathcal{B}BA^\dagger + {\rm H.c.}\right)\,.
\eeq
First, we would like to calculate the one-loop correction to the $Z$ factors. The diagrams to be considered are collected in Fig~\ref{1loop_110324}.

For diagram (a), we get
\beqn
&&\int d \theta_1 d \theta_2 \frac {d^2 q}{(2\pi)^2}\, A_1^\dagger A_2 \frac{-i}{q_L}\frac{-i}{p_L-q_L} \bar D_1 \delta(\theta_{12})\overleftarrow{D}_2\, \bar D_1 \delta(\theta_{12})\overleftarrow{D}_2\nonumber\\[3mm]
&=&\int d\theta_1\frac{d^2 q}{(2\pi)^2}\, \{D_1, \bar D_1\} A_1^\dagger\, A_1\frac 1{p_L(p_L-q_L)}\,.
\eeqn
In the above calculation we used integration by parts to move all $D$'s on one delta-function, and then do the integration over $\theta_2$. One can show that the integration over 
the momentum is finite, and, hence, this graph does not contribute to $Z_A$.

As for diagram (b), we obtain
\beqn
&&\left(i\frac\gamma2\right)^2\int d \theta_1 d \theta_2 \frac {d^2 q}{(2\pi)^2} \,B_1 B_2^\dagger \frac{-i}{p_L-q_L}\frac{-i}{q^2} \bar D_1 \delta(\theta_{12})\overleftarrow{D}_2\, \bar D_2 \delta(\theta_{12})\overleftarrow{D}_1\nonumber\\[3mm]
&=&-\left(i\frac\gamma2\right)^2\int d\theta_1 d\theta_2 \frac{d^2 q}{(2\pi)^2}\, B_2^\dagger\, B_1\frac 1{q^2(q_L-p_L)} \left( \bar D_1 D_1 \bar D_1 D_1 \delta(\theta_{12}) \right) \delta(\theta_{12})\nonumber\\[3mm]
&=& \left(i\frac\gamma2\right)^2\int d\theta_1 \frac{d^2 q}{(2\pi)^2}\, B^\dagger_1 B_1\frac{2(q_L-p_L)}{(q_L-p_L)q^2}\,.
\eeqn
Finally, it is not difficult to see that 
\beq
Z_B = 1+i\gamma^2I\,,
\eeq where the integral $I$ is defined as
\beq
I\equiv \int \frac{d^2 q}{(2\pi)^2}\frac 1{q^2}\,,
\eeq
which gives a single pole in the  UV.

Due to the fermion flavor symmetry $Z_\mathcal{B}=Z_B$, we do not need a separate calculation here. Also, at one-loop level there is no diagram contributing to $\gamma$, hence $\beta(\gamma)$ is totally determined by the $Z$ factors. In this way, we recover the result of our previous paper \cite{CS1},
\beq
\beta_{\text{one-loop}}(\gamma)=\frac {\gamma^3} {2\pi}\,.
\eeq

\begin{figure}
\begin{center}
\includegraphics[width=2in]{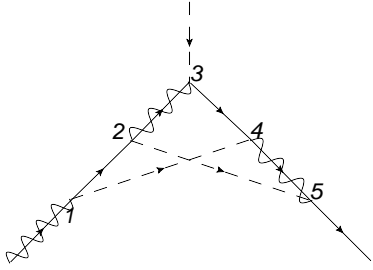}
\end{center}
\caption{\small Two-loop correction to the vertex.}
\label{gamma2l_110324}
\end{figure}

Now we are ready to move on to the
two-loop calculation. We would like to prove a version of nonrenormalization 
theorem, stating that the interaction term $\frac \gamma2 \mathcal B BA^\dagger$ is not renormalized. First, we will verify 
it at the two-loop level, by considering the diagram depicted in Fig~\ref{gamma2l_110324}. 
To this end it is sufficient  to   manipulate a little bit with the $D$-algebras,
\beqn
&&(\bar D_1 \delta(\theta_{12})\overleftarrow{D}_2)\,
(\bar D_3 \delta(\theta_{23})\overleftarrow{D}_2)\,
(\bar D_3 \delta(\theta_{34})\overleftarrow{D}_4)\,
(\bar D_5 \delta(\theta_{45})\overleftarrow{D}_4)\,
(\bar D_4 \delta(\theta_{14})\overleftarrow{D}_1)\,
(\bar D_2 \delta(\theta_{25})\overleftarrow{D}_5)\,\nonumber\\ [3mm]
&=&
\bar D_1 D_1 \delta(\theta_{12})\,\,
\bar D_3 D_3 \delta(\theta_{23})\,\,
\bar D_3 D_3 \delta(\theta_{34})\,\,
\bar D_5 D_5 \delta(\theta_{45})\,\,
\bar D_4 D_4 \delta(\theta_{14})\,\,
\bar D_2 D_2 \delta(\theta_{25})\,\nonumber\\ [3mm]
&=& - 
\bar D_1 D_1 \delta(\theta_{12})\,\,
\bar D_3\bar D_3 D_3 \delta(\theta_{23})\,\,
 D_3 \delta(\theta_{34})\,\,
\bar D_5 D_5 \delta(\theta_{45})\,\,
\bar D_4 D_4 \delta(\theta_{14})\,\,
\bar D_2 D_2 \delta(\theta_{25})\,\nonumber\\ [3mm]
&=&0\,,
\eeqn
Here we need to emphasize that the canceling is independent of the ways of regularization one takes, as we have not come to the stage of doing actual momentum integration. One can also see this explicitly from component field calculation. Q.E.D.

\begin{figure}
\begin{center}
\includegraphics[width=5.5in]{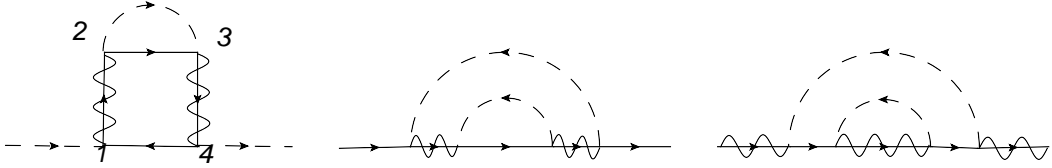}
\end{center}
\caption{\small Two-loop wave-function renormalization for $A$, $B$ and $\mathcal{B}$, respectively. 
Note that there is another diagram contributing to $Z_A$.
It gives the identical contribution to the one presented here.}
\label{A2l_110324}
\end{figure}

 With some
 extra work, one can show that due to the very same reason, 
 the two-loop correction to $Z_A$, as shown in Fig~\ref{A2l_110324}, vanishes.

There are, however, corrections to $Z_B$ and $Z_\mathcal{B}$ (see Fig.~\ref{A2l_110324}). After a straight-forward calculation we get (the subscript 0 labels the bare coupling)
\beq
Z_B = Z_\mathcal{B} = 1+i\gamma^2_0\, I+\frac12 \gamma^4_0\, I^2\,.
\eeq
The two-loop $\gamma^4$ term is an iteration  of the one-loop $\gamma^2$ term 
and has no impact on  the $\beta(\gamma)$ at the two-loop level. 
Indeed, 
\beq
\gamma^2 = \gamma^2_0 /Z_B^2\,,
\label{43d}
\eeq
and
\beq
\frac{1}{\gamma^2}  = \frac{1}{\gamma_0^2} + 2i\, I + \mbox{(possibly)}\,\, O(\gamma^4)\,,
\label{43dd}
\eeq
with no terms $O(\gamma^2)$.
The right-hand side  leads us back to $\beta(\gamma)$ as in Eq.~(\ref{simplifiedbeta}), with no two-loop contribution. 

\section{Nonrenormalization theorem in full}
\label{sec5}

Since we have both $Z_A$ and $Z_\gamma$ not corrected up to two-loop order, 
one can expect that they do not receive higher loop corrections at all. 
We will show that this is guaranteed by a nonrenormalization theorem, based on supersymmetry in conjunction with the target space symmetry of this model. Moreover, the nonrenormalization theorem is about a $D$ term rather than an
$F$ term! 

Generally speaking, each $D$ term in the Lagrangian can be treated as an $F$ term, by replacing the integration over $\theta$s by $D$s acting on the integrand. Then, following the argument of 
the $F$ term nonrenormalization, one could ask: is it
 possible to find some background 
 that preserves a half of   supersymmetry
 on which the given $F$ term does not vanish? Can one deduce, on these grounds, that a  nonrenormalization appears? 
 The answer is negative.
 
Let us first understand why  nonrenormalization theorems lose their validity for $D$ terms. Assume we want to choose a background, preserved by the supertransformation $\delta_{\bar\epsilon}$. Then, for a chiral superfield $\phi$ and its antichiral counterpart, we have
\beq
\bar D_R \phi = 0\,,\qquad  D_R \phi^\dagger = 0\,,
\label{chiralbk}
\eeq
and 
\beq
\delta_{\bar \epsilon} \phi = 0\,,\qquad \delta_{\bar \epsilon}  \phi^\dagger = 0\,,
\label{halfbk}
\eeq
where we define the supertransformation to be
\beq
\delta_{\bar \epsilon} = \bar\epsilon \frac\partial{\partial \theta_R^\dagger}+i\bar\epsilon\theta_R\partial_L\,.
\label{Qdef}
\eeq
This implies strong constraints on the background field that we could choose. Indeed, $\phi$ has to satisfy
\beq
\frac\partial{\partial \theta^\dagger_R} \phi = 0\,,\qquad \theta_R\partial_L\phi = 0\,.
\label{48}
\eeq
Equation (\ref{48}) implies, in turn a general solution of the following form: 
\beq
\phi = f(t-z,\theta_R) + g(t,z)\theta_R\,,
\eeq
where $f$ and $g$ could be arbitrary functions.
Similarly, for $\phi^\dagger$, we have
\beq
\phi^\dagger  = h(t-i\theta^\dagger_R\theta_R, z-i\theta^\dagger_R\theta_R)\,.
\eeq
Now, both $\phi$ and $\phi^\dagger$ satisfy the chiral condition. 
Therefore,  if we have a combination of $\phi$ and $\phi^\dagger$ 
and take the integral over $\int d^2 x d\theta_R d\theta^\dagger_R$, it vanishes! 

Needless to say,
if one first integrates over, say, $d\theta^\dagger_R$, and is left with the ``fake"  
$F$ term, 
the proof of the 
nonrenormalization theorem 
also fails. 
 The above $F$ term, $$\bar D_R A^\dagger \mathcal{B} B\,,$$ 
 will be a total derivative, of necessity,
and, hence, the integral over $d^2 x$ will vanish (assuming   the background to decay at infinity).

This is merely a recap of what we knew before, in a little bit fancy language. We can generalize the logic of the proof, however. In our problem  the target space symmetry reveals  itself in the invariance of the action under the   shift of $A$,
\beq
A\to A+ a(t-z)\,,\qquad A^\dagger\to A^\dagger+ a^\dagger(t-z)\,,
\label{Atranslation}
\eeq
where $a$ and $a^\dagger$ are generic functions of $t-z$. (Note that they do not need to be Hermitean-conjugate to each other.)
The reason is that the function $f(t-z)$ can be understood as 
being
both chiral and antichiral, since both $D_R$ and $\bar D_R$ vanish when acting on it. This makes it possible to combine the target space symmetry with the requirement of 
the supertransformation symmetry. Namely, we will require the background field to be invariant 
under the shift by $\delta_{\bar \epsilon}$ {\em supplemented} by the target space symmetry.

We can say that what enters in the  kinetic term for $A$ and the interaction term, 
is in fact not the field $A$ itself, but, rather, its equivalence classes under the aforementioned transformation
(\ref{Atranslation}). Let us denote by $[A]$
the equivalence class to which $A$ belongs. The key idea is that by claiming so, our constraints for the background 
field get weaker, and we have a ``thickening" of our domain
 of possible solutions to Eqs.~(\ref{chiralbk})  Eq.~(\ref{halfbk}). At the end of the day, a nontrivial background is possible.

In fact, since both the supertransformation symmetry and that of Eq.~(\ref{Atranslation}) 
are valid symmetries of $[A]$, if we pick one element in $[A]$, say, $A$, and apply $\delta_{\bar \epsilon}$, it may end up 
being another element in $[A]$, without  changing the whole equivalence class it belongs to.
Thus we can relaxe our condition (\ref{halfbk}),
\beq
\delta_{\bar\epsilon} A =\bar \epsilon a(t-z)\,,\qquad \delta_{\bar \epsilon} A^\dagger =  \bar \epsilon a^\dagger(t-z) \,,
\eeq
where $a$ and $a^\dagger$ are functions of $t-z$, and $\bar\epsilon$ is a small supertransformation parameter.

Now, this will lead us to a more general solution for the background field $A^\dagger$,
\beq
A^\dagger  =\theta^\dagger_R a^\dagger(t-z) + h(t-i\theta^\dagger_R\theta_R, z-i\theta^\dagger_R\theta_R)\,.
\eeq
Furthermore, one can also show that the allowed background for the field $A$ is not ``thickened." 
It is straightforward to verify  that by taking, for example, the following background fields:  
\beqn 
  A = g(t,z)\theta_R\,,&\qquad& A^\dagger=\theta^\dagger_R a^\dagger(t-z)\,,\nonumber\\[3mm]
 B =  1\,,&\qquad& B^\dagger=0\,,\nonumber\\[3mm]
\mathcal{B} =  f(t-z,\theta_R)\,,&\qquad&\mathcal B^\dagger = 0\,,
\eeqn
we indeed have a desirable nontrivial background for both the $A$ kinetic term   and the interaction term.

We can then apply the argumentation which leads us to the nonrenormalization theorem. 
To calculate effective action we decompose the superfields into the background and the
 quantum parts. Due to the linearity of the target space symmetry, the symmetry transformation can be assigned only
 to the background part of the $A$ field  (and, of course that of $A^\dagger$, too),  leaving the quantum part intact.
  The chosen background fields 
  are invariant under
   the transformation of $\theta^\dagger_R$ supplemented by the target space shift.
The symmetry is exact, it translates to the quantum level in form of a supersymmetry shift of $\theta^\dagger_R$. 
Therefore,
 the integrand in the loop calculations is homogeneous in $\theta^\dagger_R$, and, hence, is independent of $\theta^\dagger_R$. 
 
 On the other hand, we learn from the Feynman rules listed in Sec.~\ref{sec3} that all loop 
 calculations\,\footnote{Strictly speaking,
  this does not include the one-loop correction, since the ultraviolet contribution does not involve 
  integrations over $\theta^\dagger_R$. However, 
  the one-loop calculation is easy to carry out explicitly in the way we did it.} should involve the integration over $\theta^\dagger_R$. Thus, finally we have to obtain zero in two and higher-loop perturbative calculation. 

At the moment we are aware of no way to predict  quantum corrections for
 $\mathcal{B}$ and $B$ without explicit calculations, since the constraints  (\ref{chiralbk}) and  (\ref{halfbk}) 
 hold ``as is", leaving us with no nontrivial background for their kinetic terms. 
 
 Indeed, from 
 the loop calculation in Sec.~\ref{sec4} we can see that they get renormalized at two-loop order. Strictly speaking, the two-loop effects contain only double poles, and are merely manifestations of the one-loop terms. However, the background field method can not distinguish between a geometric progression and genuine two-loop effects. 

\section{Generalization
to nonperturbative regime a l\'a Seiberg}
\label{sec6}

In this section we will extend the 
nonrenormalization theorem  of Sec.~\ref{sec5} beyond perturbatiion theory.
We  show that $Z_A$ and $Z_\gamma$ do not receive nonperturbative corrections either. 

Following   arguments similar to that in \cite{Seiberg93}, we promote $\gamma$ to a chiral superfield. 
It is important to note that the chirality of $\gamma$ is protected by the target space symmetry. 

Indeed, let us inspect the term $\int d^2\theta_R\, \gamma\mathcal{B}BA^\dagger$. It must be invariant under the shift $A^\dagger \to A^\dagger + a^\dagger$. Then $\int d^2 \theta_R\, \gamma\mathcal{B}Ba^\dagger$ must vanish. This is impossible unless $\gamma$ is a chiral superfield. 

Now, we can assign appropriate $R$-charges to all fields. They are collected in Table~\ref{u1sym}.
\begin{table}
\begin{center}
\begin{tabular}{| c || c | c | c |}
\hline
Fields & U$(1)_1$ & U$(1)_2$ & U$(1)_3$ \\[2mm] \hline
$\gamma$ & $1$ & $0$ & $-\frac 12$ \\[2mm]
$A$ & $0$ & $1$ & $-\frac 12$ \\[2mm]
$B$ & $0$ & $1$ & $\frac 12$ \\[2mm]
$\mathcal{B}$ & $-1$ & $0$ & $-\frac 12$ \\[1mm]\hline
\end{tabular}
\end{center}
\caption{\small U(1) symmetries of the linear sigma model.}
\label{u1sym}
\end{table} 
Using these charge assignments one can show that   independent $R$-neutral combinations 
of $\gamma$, $ A^\dagger$, $B$ and $\mathcal{B}$ are
\beq
\gamma\mathcal{B}B A^\dagger\,,\quad |\mathcal{B}|^2\,,\quad |B|^2\,,\quad  |A|^2\,,\,\,{\rm and} \,\, |\gamma|^2\,.
 \eeq
Therefore,  we could   the renormalized interaction term in the effective Lagrangian in the most general case takes the form 
 \beq
 \int d^2\theta_R\, f\left(\gamma\mathcal{B}B A^\dagger,\, |A|^2,\, |\mathcal{B}|^2, \, |B|^2, \, |\gamma|^2\right)+{\rm H.c.}\,.
 \eeq
Let us suppress the dependence of $f$ on $|\mathcal{B}|^2$, $|B|^2$ and $|\gamma|^2$ for a 
short while. For a generic function of $\gamma\mathcal{B}BA^\dagger$ and $|A|^2$, it does no harm to express 
its dependence on these variables  as 
\beq
f\left(\gamma\mathcal{B}BA^\dagger, \, \frac{\gamma\mathcal{B}B}{A}\right)\,.
\eeq
 Now let us check the symmetry: under the shift symmetry $A^\dagger \to A^\dagger+a^\dagger$ for a constant $a^\dagger$, we have 
  \beqn
 &&\delta_c \int d^2\theta_R\,
  f\left(\gamma\mathcal{B}BA^\dagger, \frac{\gamma\mathcal{B}B}{A}\right)+{\rm H.c.} \nonumber\\[3mm]
 &=&\int d^2\theta_R 
 \bigg\{ \left[
 f\left(\gamma\mathcal{B}B (A^\dagger+a^\dagger), \frac{\gamma\mathcal{B}B}{A}\right) - f\left(\gamma\mathcal{B}B A^\dagger, \frac{\gamma\mathcal{B}B}{A}\right)
 \right]
 \nonumber\\[3mm]
 &&+
 \left[
 f^\dagger\left(\gamma^\dagger B^\dagger\mathcal{B}^\dagger A, \frac{\gamma^\dagger B^\dagger \mathcal{B}^\dagger}{A^\dagger +a^\dagger }\right) -  f^\dagger\left(\gamma^\dagger B^\dagger\mathcal{B}^\dagger A, \frac{\gamma^\dagger B^\dagger \mathcal{B}^\dagger}{A^\dagger }\right)
 \right] \bigg\}\,.
 \label{eq57}
 \eeqn
 The whole expression must vanish. Hence,
  we need the integrand to be a linear combination of a
  holomorphic and antiholomorphic functions.  This tells us
   that the first line must be a holomorphic function, and the second line antiholomorphic. It is straightforward to see that the former constraint requires
 \beq
 f=f_0\left(\frac{\gamma\mathcal{B}B}{A}\right)+f_1\left(\frac{\gamma\mathcal{B}B}{A}\right)\gamma
 \mathcal{B}BA^\dagger\,,
 \eeq
 where $f_{0,1}$ are some functions, generally speaking.
 In fact, $f_1$ must reduce to a constant. Otherwise, upon the shift of $A$ in its argument, we do not get a holomorphic function.  The second $D$ term in  the braces in Eq.~(\ref{eq57}) leads us to the same conclusion. 
 
 Now, let us stitch on possible dependences
  of $f_0$ and $f_1$ on $|\mathcal{B}|^2$, $|B|^2$ and $|\gamma|^2$. 
  We immediately see that they must be free of these structures. 
  
  Finally, note that the function $f_0$ will vanish under integration over $d^2\theta_R$. Hence the only term that can appear in the effective Lagrangian is $\int d^2\theta_R\, f_1\gamma\mathcal{B}BA^\dagger$. Now, since $f_1$ is independent of $\gamma$, $f_1$ has to be the canonical coefficient from the classical Lagrangian. Q.E.D.
 
 For $Z_A$ the argument is similar. Let us assume the renormalized kinetic term to be 
 \beq
 \int d^2\theta_R\, \frac12\left( fA^\dagger \partial_R A+ f ^\dagger A\partial_R A^\dagger \right)\,,
 \eeq 
 with $f$ and $f^\dagger$ generic functions of the superfields. They must be U$(1)$ neutral under the 
 $R$ rotation, according to Table~\ref{u1sym}.
 One can show, by applying the stronger symmetry,
 \beq
 A\to A+ \epsilon_1(t-z)\,,\qquad A^\dagger\to A^\dagger +\epsilon_2(t-z)\,,
 \eeq
 that the functions $f$ and $f^\dagger$ are trivial, with necessity. This completes the proof.
 
 \section{Supercurrent analysis}
 \label{sec7}
 
Here we present an alternative argument in favor of the absence of higher loops in the $\beta$ function. 

The hypercurrent we need has the form 
\beq
\mathcal{J}_{LL} = \frac 12\bar D_R A^\dagger D_R A\,.
\eeq
In components
\beq
\mathcal{J}_{LL} =j_{LL}+i\theta_R S_{LLL}
+i\theta_R^\dagger S_{LLL}^\dagger -\theta_R\theta_R^\dagger T_{LLLL}\,.
\eeq

Classically, the U$(1)_A$ current for the rotation of the chiral fermions is conserved,
\beq
j_{LL} =\psi_L^\dagger\psi_L \,,\quad \partial_R j_{LL}=0\,.
\eeq
The supercurrents are
\beq
S_{LLL} = i\sqrt{2}\partial_{L}\phi^\dagger\psi_L
\eeq
and $S_{LRR}=0$ (classically). The supercurrent concervation implies 
\beq
\partial_R S_{LRR}=0\,.
\eeq
The energy momentum tensor has the components :
\beqn
T_{LLLL} &=& -2\partial_{L}\phi^\dagger\partial_{L}\phi-i\psi_L^\dagger\partial_{L}\psi_L+i \partial_{L}\psi_L^\dagger\psi_L\,,\nonumber\\[2mm]
T_{RRRR} &=& -2\partial_{R}\phi^\dagger\partial_{R}\phi-i\psi_R^\dagger\partial_{R}\psi_R+i \partial_{R}\psi_R^\dagger\psi_R-i\zeta_R^\dagger\partial_{R}\zeta_R\\[2mm]
&&+i \partial_{R}\zeta_R^\dagger\zeta_R-2[i\gamma\zeta_R\psi_R\partial_{R}\phi^\dagger+\text{H.c.}]\,,\nonumber\\[2mm]
T_{LLRR}&=&0\,\text{   (classically) }.\nonumber
\eeqn
It is easy to see that the three currents $j_{LL}$, $S_{LLL}$ and $T_{LLLL}$ form a $\mathcal{N}=(0,2)$ (nonchiral) supermultiplet, which we denote by $\mathcal{J}_{LL}$ and refer to as the hypercurrrent.
In superfields we can write $\partial_R \mathcal{J}_{LL}=0$.

Quantum mechanically $j_{LL}$ is no longer conserved, due to the chiral fermion anomaly, and hence the conservation laws are adjusted in terms of superfields
\beq
\mathcal{W}_R = -\frac{i\gamma^2}{4\pi}\bar D_R(B^\dagger B+\mathcal{B}^\dagger \mathcal{B})\,,
\eeq
which, in component, is
\beq
\mathcal{W}_R=-S_{LRR}^\dagger +i\theta_R(T_{LLRR}+i\partial_{R}j_{LL})+i\theta_R\theta_R^\dagger \partial_{L}S_{LRR}^\dagger
\eeq
In particular, there will be a nontrivial contribution to $S_{LRR}$ and $T_{LLRR}$:
\beqn
&&S_{LRR}=- \frac i{\sqrt{2}\pi}\gamma^3\psi_L\psi_R^\dagger\zeta_R^\dagger\,,\nonumber\\
&&T_{LLRR}=-\frac{ \gamma^2}{2\pi}\left[\gamma^2\psi_L^\dagger\psi_L(\psi_R^\dagger\psi_R+\zeta_R^\dagger\zeta_R)-i\psi_R^\dagger\overset{\leftrightarrow}{\partial_{L}}\psi_R-i\zeta_R^\dagger\overset{\leftrightarrow}{\partial_{L}}\zeta_R\right]\,.
\eeqn

\begin{figure}
\begin{center}
\includegraphics[width=3.5in]{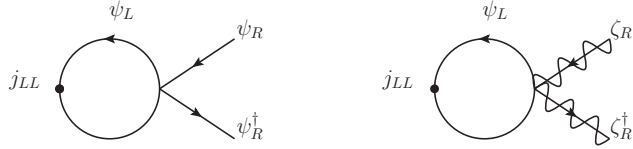}
\end{center}
\caption{\small One-loop diagram for $j_{LL}$ anomaly. }
\label{j5ano_110922}
\end{figure}

Thus the chiral anomaly (see Fig.~\ref{j5ano_110922}) and supersymmetry fix the trace of the energy momentum $T^\mu_\mu$, which is proportional to the $\beta$ function. Moreover, we could absorb the power of $\gamma$ into the definition of the fields, which means that
\beq
T_{LLRR} = \frac{2}{\gamma}\beta(\gamma)\mathcal{L}\,.
\eeq 

From this we can see that the $\partial_R j_{LL}$ anomaly actually controls the running of the coupling of the theory. Since the chiral fermion anomaly is a one-loop effect, there is no higher loop contribution to $\partial_{R}j_{LL}$, which also implies that the $\beta$ function of $\gamma$ is one-loop exact. Recall that $\beta$ function also encodes the information of wave-function renormalization of $\zeta_R$ and $\psi_R$, we could indirectly show that their anomalous dimensions are also one loop exact. This will be elaborated in more detail in the subsequent publication \cite{CS3}. 
\section{Conclusion}

In this paper, we introduce a simplified but instructive
model that illustrates the nature of the heterotic deformation of $\mathcal{N}=(2,2)$ to $\mathcal{N}=(0,2)$ theories. 
It was that the theory should have some conformal properties, see e.g. \cite{SY1}.  We  showed that this is partially true, due to the nonrenormalization of the interaction term and the target field $A$. The supergraph method 
for  the $\mathcal{N}=(0,2)$ case that we worked out prompted us that we should expect some nonrenormalization 
theorems.
This is due to the fact that   relevant diagrams vanish at the level of the
$D$-algebra --- before the momentum integration. And indeed, the nonrenormalization theorems did materialize!

The most interesting result is the proof of $D$ term nonrenormalization for the $A$ kinetic and interaction terms. 
We  generalized the conventional  procedure and demonstrated
 that invoking the target space symmetries 
 we can in a sense  expand in realm of $F$ terms. 
The key fact is that the target space symmetry ``thickens" the solution for the nontrivial background field. 
Actually this has a deep relation to the equivariant $Q$-cohomology classes, which 
may provide us with a new standpoint for  generalization of some of the above arguments to certain models, e.g., 
the heterotic CP$(N-1)$ models. We will continue to study the 
nonlinear version of this result in our forthcoming paper \cite{CS3}.

\section*{Acknowledgments}

XC thank T. Lawson for inspiration in an important stage of this research. We are grateful to J. Chen and
T. Dumitrescu for illuminating discussions.

XC is supported in part by the Hoff Lu Fellowship in Physics at the University of Minnesota. The work of MS is supported in part by DOE grant DE-FG02-94ER408.


\end{document}